\documentclass[epj]{svjour}%,referee
\usepackage{graphicx}
\usepackage{amsmath}
\usepackage{amssymb}
\begin{document}
\title{Quantum synchronization}
%\subtitle{Quantum synchronization}
\author{O. V. Zhirov\inst{1} \and D. L.~Shepelyansky\inst{2}% etc
% \thanks is optional - remove next line if not needed
%\thanks{\emph{Present address:} Insert the address here if needed}%
}                     % Do not remove
%
%\offprints{}          % Insert a name or remove this line
%
\institute{ Budker Institute of Nuclear Physics, 630090 Novosibirsk, Russia
\and 
Laboratoire de Physique Th\'eorique, 
UMR 5152 du CNRS, Univ. P. Sabatier, 31062 Toulouse Cedex 4, France}
\date{Received: date / Revised version: date}
% The correct dates will be entered by Springer
%
\abstract{
Using the methods of quantum trajectories 
we study numerically the phenomenon of quantum
synchronization in a quantum dissipative
system with periodic driving.
Our results show that at small values
of Planck constant $\hbar$ the classical devil's staircase
remains robust with respect to quantum fluctuations
while at large $\hbar$ values synchronization
plateaus are destroyed. Quantum synchronization in our model
has close similarities with Shapiro steps in Josephson junctions
and it can be also realized in experiments with cold atoms.
 % end of PACS codes
} %end of abstract
\maketitle
Since 1665, when Christiaan Huygens discovered the synchronization
of two maritime pendulum clocks \cite{huygens} 
(see \cite{weisenfeld,pikovsky} for historical survey and modern experiments), 
it has been established that this universal nonlinear phenomenon
appears in an abundant variety of systems ranging from
clocks to fireflies, cardiac pacemakers, lasers and
Josephson junction (JJ) arrays \cite{pikovsky}.
Various mathematical tools have been developed
to analyze synchronization in simple 
dissipative nonlinear models with periodic driving
and complex ensembles of nonlinear coupled oscillators.
Such pure mathematical concepts as Arnold tongues in
the circle map \cite{arnold} found their
experimental implementations with Shapiro steps
\cite{shapiro} and JJs synchronization \cite{likharev}.
However, till recently  the treatment of synchronization
has been mainly based on classical mechanics
even if JJs have purely quantum origin \cite{pikovsky,likharev}.

A significant progress  in the theory of dissipative quantum mechanics
has been done in Ref.\cite{leggett}. It was further
developed by various groups as reviewed in \cite{weiss}.
Nowadays this research line is getting more and more importance
since technology goes on smaller and smaller
scales where an interplay of dissipative
and quantum effects becomes dominant. A typical example
is given by small size JJs. Here dissipative
effects are always present even if in certain cases
skillful manipulations allow to realize long term
coherent Rabi oscillations \cite{esteve}.

These reasons led to a significant number of
analytical studies of quantum tunneling in
presence of dissipation (see e.g. 
\cite{leggett,weiss,leggett1,ovchinnikov,averin}).
However, analytical methods have serious restrictions
in a strongly nonlinear regime typical of
synchronization \cite{pikovsky}. Due to that
in this paper we perform extensive numerical
studies of quantum synchronization following
precisely a transition from 
classical to quantum behavior changing
effective dimensionless Planck constant $\hbar$
by three orders of magnitude.
The quantum dissipative evolution is described by the
master equation for the density matrix $\rho$
written in the Lindblad form \cite{lindblad}.
To perform simulations with a large
number of states $N \propto 1/\hbar$ 
we use the method of quantum trajectories
\cite{schack,brun} which allows to reduce
significantly the number of equations
compared to direct solution of the
master equation ($N$ instead of $N^2$).
In this approach one quantum trajectory 
can be viewed as one individual realization
of experimental run \cite{dalibard}.

For our studies we choose a model
which in the classical limit
is closely related to the circle map \cite{arnold}
which gives a generic description
of Arnold tongues and synchronization of
system dynamics with an external periodic driving
\cite{pikovsky}. In absence of dissipation
the evolution is described by the Hamiltonian of kicked
particle falling in a static field $f$:
\begin{equation}
\hat{H}=\hat{p}^2/2 \; - f \hat{x} +
K\cos\, \hat{x} \sum_{m=-\infty}^{+\infty}\delta(t-m),
\label{eq1}
\end{equation}
with the usual operators $\hat{x}$
and $\hat{p} = \hbar \hat{n}=-i \hbar d/dx$.
At $f=0$ this Hamiltonian corresponds to the kicked rotator
\cite{izrailev}, a paradigmatic model in the fields of nonlinear 
dynamics and quantum chaos. At  non-zero $f$ the
system has been built up in experiments
with cold atoms placed in laser fields and
falling in a gravitational field  \cite{darcy}.
Nontrivial quantum effects of the static force
are analyzed in \cite{fishman}. Here we consider
the dynamics of the model in presence of
additional friction force $F = -g^2 p$.
Experimentally such a force can be realized
by the Doppler cooling \cite{scully}.
The model can be also implemented with JJs 
as described in \cite{graham}.
In this case effective  kicks are
created by an external {\it ac-}current
source, $f$ and $p$ are proportional respectively
to {\it dc-}current and voltage drop across JJ,
while the friction force $F$
naturally appears due to finite circuit resistance.
The expression of $K$ and $\hbar$ via JJ parameters
is given in \cite{graham}.

\begin{figure}[th]
   \centering
   \includegraphics[width=0.48\textwidth,angle=0]{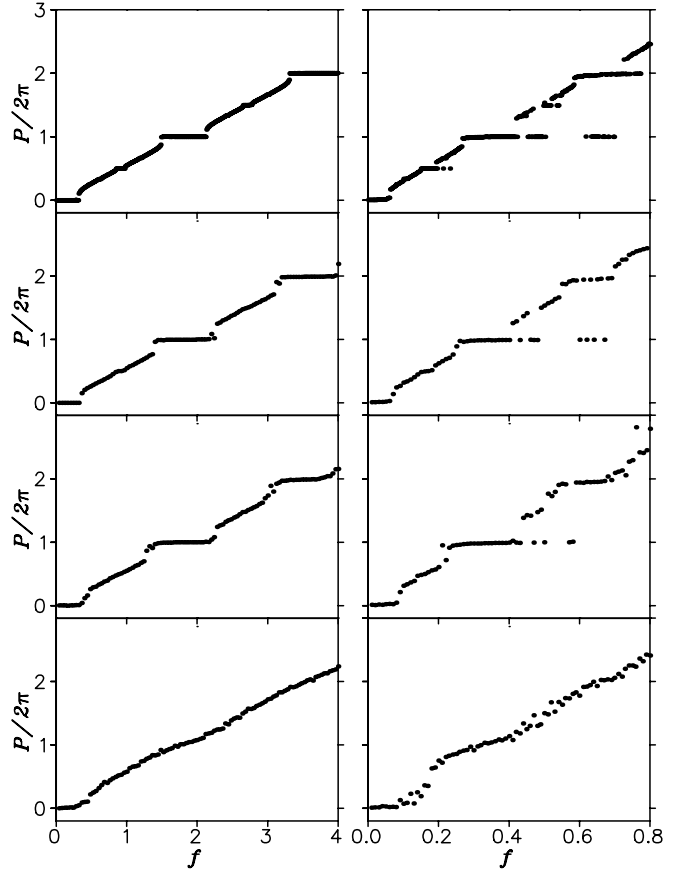}
   \vglue -0.00cm
   \caption{Dependence of the average momentum $P$ 
on static force $f$ at $K=0.8$ for
$\gamma=0.25$ (left column) and $\gamma=0.05$ (right column);
$P$ is computed over $t= 500$ map iterations.
From top to bottom: classical case at $\hbar=0$,
$\hbar=0.012$,  $\hbar=0.05$, $\hbar=0.5$.
Initial conditions correspond to 
one classical trajectory at  $x=0$, $p/2\pi=0.38$
for classical dynamics. For the quantum evolution
one quantum trajectory is taken at the same 
$x,p$ position with the wave function in the form
of minimal coherent state at given $\hbar$.
    }
   \label{fig1}
\end{figure}

The classical dynamics (\ref{eq1}) with friction
can be exactly integrated between kicks
that gives a dissipative map
\begin{equation}
\begin{array}{l}
\bar{p} = (1 - \gamma) p + (1-\gamma)K \sin \, x + f \gamma/g^2 \, , \\
\bar{x} = x + \gamma p /g^2 + 
(\gamma K/g^2) \sin \, x + f(g^2-\gamma)/g^4 \, ,
\end{array}
\label{eq2}
\end{equation}
where bars note new values of variables after one map iteration
and $1-\gamma = \exp(-g^2)$. Up to parameter rescaling and
shifts in $x, p$, produced by static force, the map (\ref{eq2})
has the form of Zaslavsky map \cite{zaslavsky}.
Due to contraction in $p$ the dynamics
in phase variable $x$ is close to
the circle map $\bar{x} = x + K_{eff} \sin \, x + \nu$ 
\cite{pikovsky,arnold}
and demonstrates devil's staircase structure 
in the dependence of average momentum $P$ on $f$
(Fig.~\ref{fig1}, top). Steps near rational rotation numbers
$P/2\pi$ correspond to synchronization
with external  periodic driving inside Arnold tongues.
In average the momentum $P = f/g^2$
as it should be in an equilibrium between  
the external and friction forces. 

The corresponding quantum dissipative dynamics 
is described by the master equation
in the Lindbald form \cite{lindblad}:
\begin{equation}
\dot{\hat{\rho}} = -i
[\hat{H},\hat{\rho}] - \frac{1}{2} \sum_{\mu}
\{\hat{L}_{\mu}^{\dag} \hat{L}_{\mu},\hat{\rho}\}+
\sum_{\mu} \hat{L}_{\mu} \hat{\rho} \hat{L}_{\mu}^{\dag},
\label{eq3}
\end{equation}
where $\hat{\rho}$ is the density operator,
\{\,,\,\} denotes the anticommutator, 
$\hat{L}_{\mu}$ are the Lindblad operators, 
which model the effects of the environment.
Following \cite{gabriel} we assume the Lindblad operators
in the form ($\mu=1,2$):
\begin{equation}
\begin{array}{l}
\hat{L}_1 = g \sum_n \sqrt{n+1} \; |n \rangle \, \langle n+1|,\\
\hat{L}_2 = g \sum_n \sqrt{n+1} \; |-n \rangle \, \langle -n-1| .
\end{array}
\label{eq4}
\end{equation}
These operators act on the bases of $2\pi$-periodic eigenstates
of operator $\hat{n}$ and in the regime of weak coupling and
Markov approximations describe the dissipation force
$F=-g^2 p$ induced by a bosonic bath at zero temperature.
As in \cite{gabriel} the numerical simulations of quantum jumps
are done for one quantum trajectory using the so-called
Monte Carlo wave function approach \cite{dalibard}.
The additional term with the constant force $f$
is exactly integrated between jumps leading to a drift 
of wave function amplitudes in the space of momentum
eigenstates  $n$. The total number of states
$N$ is fixed by a condition of keeping all states
with probabilities higher than $10^{-7}$.

The numerical results for the quantum devil's staircase
are shown in Fig.~\ref{fig1} for various $\hbar$ values \cite{note0}.
The dependence is similar to those of $I - V$ characteristics
shown in \cite{shapiro} with $P \propto V$, $f \propto I$.
At small $\hbar$ values ($\hbar=0.012$) the steps 
remains stable with respect to quantum fluctuations.
As a result, the rotation frequency
is unchanged when $f$ is varied in some finite interval.
This can be viewed as 
the manifestation of 
quantum synchronization which takes place inside
quantum Arnold tongues.
For larger $\hbar = 0.05$
small steps start to disappear and at 
relatively large $\hbar=0.5$ the dependence
$f - P$ becomes smooth so that a quantum particle
slides smoothly under static force $f$.
It is interesting to compare the cases of strong
$\gamma=0.25$ and weak $\gamma=0.05$
dissipation (Fig.~\ref{fig1} left and right columns).
At strong $\gamma$ the contraction in $p$
is rather rapid and the behavior is similar
to the case of the circle map \cite{arnold}
and $P$ grows monotonously with $f$.
At weak $\gamma$ there are many different attractors
in the phase space and a classical trajectory
may jump between them rather irregularly
with variation of $f$ or initial conditions.
At small $\hbar$ a quantum trajectory reproduces this behavior 
of steps overlap rather accurately. We note that
very similar structure of steps overlap
is clearly seen in experimental data shown
in Fig.2 of \cite{shapiro}.
\begin{figure}[th]
   \centering
   \includegraphics[height=0.48\textwidth,angle=90]{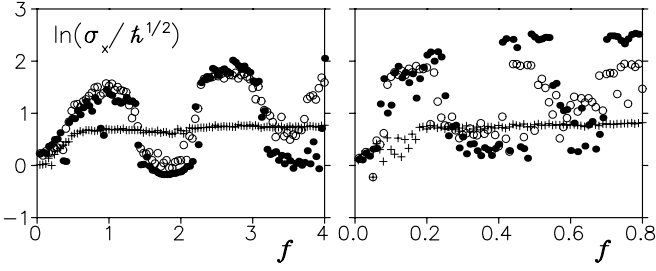}
   \vglue -0.00cm
   \caption{Dependence of rescaled
dispersion  $\sigma_x/\hbar^{1/2}$ of wave packet in $x$
on force $f$ for the cases of Fig.1
at $\gamma=0.25$ (left panel) and $\gamma=0.05$ (right panel).
Symbols show data at
$\hbar=0.012$ (black points), $\hbar=0.05$ (open circles),
$\hbar=0.5$ ($+$).
Dispersion $\sigma_x$ is averaged over $t=500$ 
map iterations for one quantum trajectory
with initial state of Fig.~\ref{fig1}.
%$|\psi(x)|^2 \sim \exp(-x^2/(2 \sigma_x^2))$.
    }
   \label{fig2}
\end{figure}

To understand the properties of dissipative quantum dynamics
we analyze the structure of wave functions associated
with quantum trajectories of Fig.\ref{fig1}. It is known that
dissipation in quantum systems leads 
to a collapse of wave packet
\cite{percival,linear,schack1}. In agreement with these results
and recent studies of Zaslavsky map \cite{gabriel}
we find that  the wave function in our model collapses on
a compact  packet of width $\sigma_x$, $\sigma_p$
(dispersion)
in coordinate and momentum respectively
(e.g. $|\psi(x)| \sim  \exp(-x^2/(4 \sigma_x^2))$)
\cite{note}. The width $\sigma_x$ depends nontrivially
on $f$ as it is shown in Fig.\ref{fig2}. Inside the steps
with synchronization 
the value of $\sigma_x$ is strongly reduced
while in the sliding regime between steps
its value is significantly enhanced.
In the limit of small $\hbar$ where quantum synchronization
takes place the rescaled width $\sigma_x/\sqrt{\hbar}$
shows approximately the same functional dependence on $f$
for all $\hbar$. This functional dependence
on $f$ is completely changed at large
$\hbar$ where synchronization is absent
and where $\sigma_x$ is not very sensitive to $f$.
For small $\gamma$ (Fig.\ref{fig2} right)
the dependence on $f$ becomes more irregular
in agreement with more complicated step
structure seen in Fig.\ref{fig1} (right).
We note that at the same time $\sigma_p$
remains practically insensitive to variation of $f$. 
This is related to the fact that 
in momentum $p$ the dynamics is close
to a simple contraction while in coordinate
$x$ the nonlinear dynamics significantly depends on the
system parameters (see discussion below).
\begin{figure}[th]
   \centering
   \includegraphics[height=0.48\textwidth,angle=90]{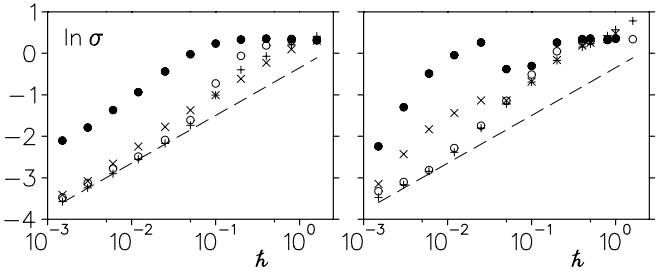}
   \vglue -0.00cm
   \caption{Dependence of dispersion $\sigma$ in $x$ ($\sigma_x$)
   and $p$ ($\sigma_p$) on $\hbar$ for
   $K=0.8$ at $\gamma=0.25$ (left) and 
   $\gamma=0.05$ (right); $\sigma$ is averaged over $t=5000$
   map iterations.
   Left panel: $f=1.1$ (points for $\sigma_x$ and $\times$ for $\sigma_p$),
   $f=1.9$ (open circles for $\sigma_x$ and $+$ for $\sigma_p$ ).
   Right panel: $f=0.21$ (points for $\sigma_x$ and $\times$ for $\sigma_p$),
   $f=0.37$ (open circles for $\sigma_x$ and $+$ for $\sigma_p$ ).
   Dashed line shows the dependence $\sigma=(\hbar/2)^{1/2}$.
   Initial state is as in Fig.~\ref{fig1}.
    }
   \label{fig3}
\end{figure}

\begin{figure}[th]
   \centering
   \includegraphics[width=0.48\textwidth,angle=0]{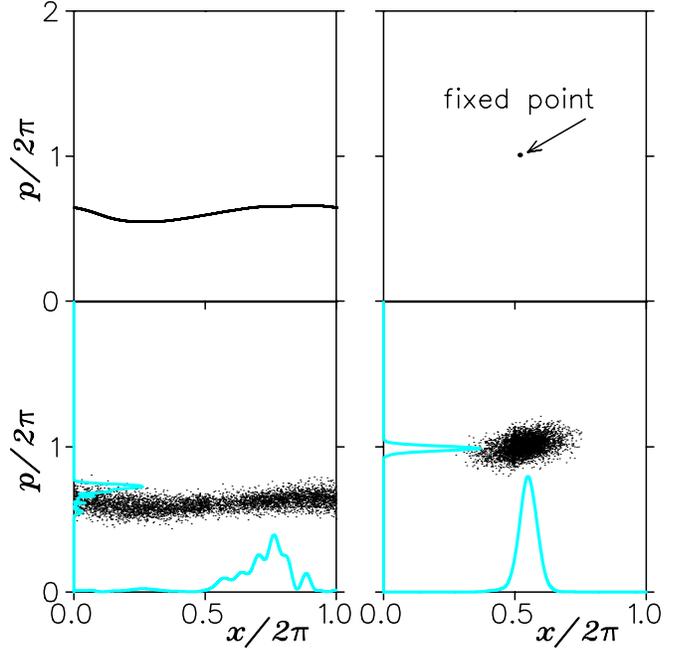}
   \vglue -0.00cm
   \caption{(Color online) Poincar\'e section for $K=0.8$, $\gamma=0.25$ at
    $f=1.1$ (left) and  $f=1.9$ (right).
    Top: classical case; bottom: quantum case at $\hbar=0.05$,
    points mark the average $x,p$ positions of wave packet;
    light blue (gray) curves on $x,p$ axes show at $t=5000$ the
    quantum probability distribution in $x$ and $p$ respectively
    (arbitrary units).    
    Number of map iterations $t=5000$, 
    initial state is as in Fig.~\ref{fig1}. }
   \label{fig4}
\end{figure}

To check more accurately the scaling $\sigma_x, \sigma_p \propto \sqrt{\hbar}$
we fix two values of force $f$ 
and vary $\hbar$ by three orders of magnitude (see Fig.\ref{fig3}).
One value of $f$ is taken on a synchronization plateau
when a classical attractor is a fixed point in the phase
space and another is taken on a slope
when the classical dynamics has attractor in a form of invariant
curve as it is shown in Fig.~\ref{fig4} for $\gamma=0.25$.
Clearly $\sigma_x$ is significantly larger
in the case of invariant curve than in the case of fixed point.
Intuitively it is rather natural since in the first case
variation of phase $x$ is unbounded in $2\pi$
while in the later case the phase value is fixed (Fig.~\ref{fig4}).
Thus we may conclude that quantum synchronization
gives a significant reduction of quantum fluctuations. 
At the same time fluctuations in $p$ characterized by $\sigma_p$
are not sensitive to the choice of $f$.
In the limit of small $\hbar$ the numerical data
of Fig.\ref{fig3} clearly show that the wave packet 
dispersion scales as 
\begin{equation}
\sigma_x \sim \sigma_p \propto \sqrt{\hbar} \; .
\label{eq5}
\end{equation}
%\vglue -0.30cm
In the synchronization regime 
the dispersion is even close to its minimum
value with $\sigma_x \sigma_p = \sigma_p^2=\hbar/2$
(dashed line in Fig.\ref{fig3}). At smaller dissipation
$\gamma$ the asymptotic
dependence $\sigma \propto \sqrt{\hbar}$
starts at smaller values of $\hbar$ (cf. left and right panels of
 Fig.\ref{fig3}). Indeed, 
the wave packet width grows dispersively 
during time $1/\gamma$ that makes $\sigma$ larger 
at small $\gamma$. The dependence (\ref{eq5}) is in agreement
with the data obtained in the regime of wave packet collapse
in \cite{gabriel} for a smaller
range of $\hbar$ variation  (note also \cite{note}).
\vglue -0.00cm
Other properties of quantum dissipative dynamics
can be understood from Poincar\'e phase space plots.
In the classical case two typical examples
are shown in Fig.\ref{fig4} (top) with invariant curve
and fixed point attractors. In the corresponding quantum case
we plot in the phase space the values of  $x$, $p$ 
averaged over a wave function of a
given quantum trajectory  at each map iteration
(integer time $t$) (Fig.\ref{fig4} bottom,
quantum probability distributions are also shown).
The quantum  Poincar\'e plot reproduces correctly
the global structure of classical phase space
but additional noise produced by quantum fluctuations
is clearly visible. This quantum noise
creates an effective width of an invariant curve
and replaces a fixed point by a spot of finite size.
The size of these fluctuations is approximately given
by $\sigma_x$ and $\sigma_p$.
They  decrease
with $\hbar$  according to the relation (\ref{eq5}).
Due to this quantum noise two quantum trajectories
with the same initial state can drop on different attractors
and thus contribute to different plateaus
in the devil's staircase (Fig.\ref{fig1}). 
We indeed observed such cases simulating different quantum
trajectories with the same initial state.
However, convergence to different plateaus 
takes place only at relatively weak dissipation
($\gamma=0.05$) when there are many different
classical attractors close in a phase space
and quantum noise may give transitions between them.
At strong dissipation ($\gamma=0.25$) the
attractors are well separated and different quantum trajectories
converge to the same plateau.

\begin{figure}[th]
   \centering
   \includegraphics[width=0.22\textwidth,angle=90]{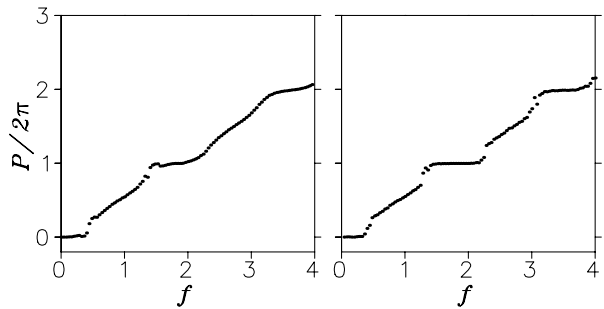}
   \vglue -0.00cm
   \caption{Dependence of the average
   momentum $P$ on static force $f$ at $K=0.8$,
   $\gamma=0.25$ (compare with the 
   data in Fig.~\ref{fig1}, left column). Right: quantum case at 
   $\hbar = 0.05$ (same data as in Fig.~\ref{fig1}, left column).
   Left: classical simulation of classical dynamics
   with 200 classical trajectories
   in presence of noise in $x,p$ with the dispersion of noise
   corresponding to the values
   $\sigma_x$ and $\sigma_p$ taken from the quantum 
   trajectory simulation at $\hbar=0.05$ (right).
    }
   \label{fig5}
\end{figure}

The above  results show that the quantum synchronization
has close links with the classical synchronization
in presence of noise. The later has been intensively
investigated and it is known that the synchronization plateaus
are preserved if the noise amplitude is significantly smaller
than their size \cite{pikovsky}. In a similar way
we may conclude that the quantum synchronization
remains robust with respect to quantum fluctuations
until their amplitude 
(proportional to $\sigma_x, \sigma_p \propto \sqrt{\hbar}$)
remains smaller than the size of a synchronization plateau.
We also checked that the classical dynamics (\ref{eq2})
with additional noise in $x,p$, with dispersion 
values $\sigma_x, \sigma_p$ taken from the quantum evolution,
gives the dependence $P$ vs. $f$  which is close to the
quantum result. This is illustrated in Fig.~\ref{fig5}.
This correspondence
remains valid if the size of the wave packet
is sufficiently small ($\sigma_x, \sigma_p \ll 1$).
However,  at large $\hbar$
or small $\gamma$ when $\sigma_x, \sigma_p \sim 1$
this correspondence is broken and
the quantum fluctuations cannot be reduced to
effects of classical noise.

Above we discussed quantum synchronization with
external periodic driving. It would be
interesting to investigate the synchronization
of two quantum nonlinear pendulum clocks 
to have a quantum version of the Huygens experiment \cite{huygens}.
Recently, first numerical simulations in 
such kind of regime (two quantum coupled Duffing oscillators)
has been performed in \cite{everitt}. Their results
show that in the case when the quasi-integrable
dynamics of oscillators becomes synchronized
(entrained) the von Neumann entropy of
one oscillator drops significantly.
This result is in a qualitative agreement
with our data showing that the dispersion of quantum
state drops significantly on synchronization plateaus
(Fig.~\ref{fig2}) \cite{note1}. Further investigations
of quantum synchronization in coupled
nonlinear systems are of great interest.
For example, quantization may produce
nontrivial effects in the synchronization
transition in a disordered JJ series array discussed in 
\cite{wiesenfeld1}.

Modern technological progress allows to
study the regime of quantum synchronization with 
small size JJs (e.g. similar to those of \cite{esteve})
or with cold or BEC atoms (e.g. like in \cite{darcy}).
This should open new perspectives for synchronization
of quantum objects with possible applications
to quantum computations.

This work was supported in part by the EC IST-FET projects EDIQIP,
EuroSQIP and (for OVZ) by RAS Joint scientific program 
"Nonlinear dynamics and solitons".

\vglue -0.30cm


\begin{thebibliography}{10}
\bibitem{note0} We show here data for one typical case at $K=0.8$,
     similar results have been seen in the range $0.4 \leq K \leq 2$
     with quasi-integrable dynamics.
\bibitem{note} We study here the nonlinear dynamics
     in a quasi-integrable regime without Ehrenfest explosion
     \cite{gabriel}.
\bibitem{note1} Let us note that a large 
     value of entropy found in \cite{everitt}
     in the chaotic regime at small $\hbar$ should be related to the
     Ehrenfest explosion discussed in \cite{gabriel}.
\end{thebibliography}
\end{document}